\documentclass[superscriptaddress]{revtex4}
\usepackage{graphicx} 

\begin{document}
\title{Angle-dependence of quantum oscillations in  YBa$_\mathbf{2}$Cu$_\mathbf{3}$O$_\mathbf{6.59}$
shows free spin behaviour of quasiparticles}


\author{B.~J.~Ramshaw}
\affiliation{Department of Physics and Astronomy, University of British
Columbia, Vancouver, BC, V6T~1Z1, Canada}
\author{Baptiste Vignolle}
\affiliation{ Laboratoire National des Champs Magn\'etiques Intenses, UPR 3228, CNRS-INSA-UJF-UPS, Toulouse, France}
\author{James Day}
\affiliation{Department of Physics and Astronomy, University of British
Columbia, Vancouver, BC, V6T~1Z1, Canada}
\author{Ruixing Liang}
\affiliation{Department of Physics and Astronomy, University of British
Columbia, Vancouver, BC, V6T~1Z1, Canada}
\affiliation{Canadian Institute for Advanced Research, Toronto, Canada}
\author{W.N. Hardy}
\affiliation{Department of Physics and Astronomy, University of British
Columbia, Vancouver, BC, V6T~1Z1, Canada}
\affiliation{Canadian Institute for Advanced Research, Toronto, Canada}
\author{Cyril Proust}
\affiliation{ Laboratoire National des Champs Magn\'etiques Intenses, UPR 3228, CNRS-INSA-UJF-UPS, Toulouse, France}
\author{D.~A.~Bonn}
\affiliation{Department of Physics and Astronomy, University of British
Columbia, Vancouver, BC, V6T~1Z1, Canada}
\affiliation{Canadian Institute for Advanced Research, Toronto, Canada}

\begin{abstract}
Measurements of quantum oscillations in the cuprate superconductors afford a new opportunity
to assess the extent to which the electronic properties of these materials yield to a
description rooted in Fermi liquid theory. However, such an analysis is hampered by the small
number of oscillatory periods observed. Here we employ a genetic algorithm to globally model
the field, angular, and temperature dependence of the quantum oscillations observed in the
resistivity of $YBa_2Cu_3O_{6.59}$. This approach successfully fits an entire data set to a
Fermi surface comprised of two small, quasi-2-dimensional cylinders. A key feature of the
data is the first identification of the effect of Zeeman splitting, which separates spin-up
and spin-down contributions, indicating that the quasiparticles in the cuprates behave as
nearly free spins, constraining the source of the Fermi surface reconstruction to something
other than a conventional spin density wave with moments parallel to the $CuO_2$ planes.
\end{abstract}

\maketitle

A magnetic field H applied to a metal puts electrons into discrete Landau levels whose
spacing increases linearly with the field. As a consequence, the physical properties of a
metal oscillate with a characteristic 1/H periodicity due to successive Landau levels passing
through the Fermi energy which separates filled from empty states. Following the initial
observation of quantum oscillations in the high temperature superconductor
YBa$_2$Cu$_3$O$_{6+x}$ via Hall effect and in-plane resistivity
measurements\cite{Doiron:2007}, there have been reports of similar effects in
magnetization\cite{Jaudet:2008}, and tunnel diode oscillator
measurements\cite{Sebastian:2008}. While initially seen in high purity oxygen-ordered
YBa$_2$Cu$_3$O$_{6.51}$ and in the stoichiometric compound
YBa$_2$Cu$_4$O$_8$\cite{Bangura:2008,Yelland:2008}, quantum oscillations have also been
observed in both Tl$_2$Ba$_2$CuO$_{6+\delta}$\cite{Vignolle:2008} and in the electron doped
cuprates\cite{Helm:2009} in spite of their greater cation disorder. Comparisons between
different compounds and dopings suggest a drastic reorganization of the electronic structure,
from a large hole-like Fermi surface in overdoped compounds\cite{Vignolle:2008} to small
pockets in underdoped compounds. Recently, measurements in the YBa$_2$Cu$_3$O$_{6+x}$ (YBCO)
system have been exploring the underdoped side in greater detail, to determine the extent to
which the measurements can be modeled by conventional Fermi liquid treatments, despite the
strong electronic correlations in the underdoped region. In these studies, additional
oscillation frequencies and thus a more complicated Fermi surface have been
reported\cite{Sebastian:2008,Audouard:2009, Sebastian:2010}. We concentrate here on the
observation that the original low frequency oscillation is actually comprised of multiple,
closely spaced frequencies\cite{Audouard:2009}. In this situation, angle dependent
measurements clarify whether the multiple frequencies arise from a warped quasi-2D Fermi
cylinder. Below, we present such a set of measurements together with an analysis method that
uncovers for the first time the effects of Zeeman splitting.

In Fig. 1 we present oscillations of the longitudinal resistivity (Shubnikov-de Haas effect)
in the direction perpendicular to the CuO$_2$ planes ($\hat{c}$-axis), and with the magnetic
field also aligned along the $\hat{c}$-axis. Although it is the $\hat{c}$-axis resistivity
being measured, the oscillations are probing the cyclotron motion of electrons within the
CuO$_2$ planes. At 1.5 K, the onset of non-zero resistivity shows the vortex lattice begins
to depin or melt at 24 Tesla. This transition occurs at lower fields with increasing
temperature, giving a trade-off between available field range and oscillation amplitude. The
derivative of these curves reveals that the oscillations begin at the onset of the melting
transition, but we concentrate on the regime well above depinning, where the background shape
of the resistivity is simpler.

A challenge in analyzing data from the underdoped cuprates is that the oscillation frequency
is low and there is a limited field range available. Even if one could reach much higher
fields, the number of extra periods would not be large, since at 60 Tesla the samples are
already in the regime where fewer than ten Landau levels remain. By way of comparison, de
Haas-van Alphen measurements in the ruthenates have thousands of periods of oscillation
between 2 and 33 Tesla, enabling detailed analysis of several frequencies arising from three
or more distinct, cylindrical Fermi surfaces, each with high-order warping terms due to
$\hat{c}$-axis coupling\cite{Bergemann:2003}. In the case of the ruthenates, Fourier
transforms and windowing techniques could be used to isolate individual frequency components
and to analyze their amplitude dependencies on angle, field and temperature according to
Lifschitz-Kosevich theory \cite{Bergemann:2002}. For underdoped YBa$_2$Cu$_3$O$_{6+x}$, a
slight minimum can be seen in the oscillation amplitude near 47 Tesla (see inset of figure
1). This ``beat'' indicates that there are two or more closely-spaced frequencies, but the
small number of periods means that they would be hard to distinguish as individual
frequencies using standard Fourier transform methods\cite{Audouard:2009}.

One explanation for the close frequencies indicated by the beats in the data is a cylindrical
Fermi surface with a small warping. Roughly speaking, warping manifests itself as a splitting
corresponding to ``neck'' and ``belly'' frequencies, associated with the minimum and maximum
diameter of a Fermi surface cylinder along the $\hat{c}$-axis.  If the splitting is due to
warping, the frequency components ought not to be treated separately but should be dealt with
using the integral expressions for Landau levels passing through such a surface. For a single
cylindrical pocket with sinusoidal warping, the expression for the normalized resistivity of
the fundamental component of oscillation is
\begin{equation}\label{osc}\rho_{osc} / \rho_o = A R_T R_{D} R_s \sin \left( 2 \pi \left( \frac{F}{H \cos\theta} - \gamma \right) \right) J_0\left(2 \pi
\frac{\Delta F} {H \cos\theta} J_0 \left( k_F c \tan\theta \right)\right),
\end{equation}
where $F$ is to the average area of a cylinder, $\Delta F$ is the splitting due to warping,
$k_F$ is the average radius of the cylinder, $c$ is the $\hat{c}$-axis lattice parameter,
$\theta$ is the angle between the field and the $\hat{c}$-axis, $\gamma$ is a phase factor,
$R_T$, $R_D$, and $R_s$ are amplitude reduction factors discussed below, and $A$ is an
overall amplitude\cite{Yamaji:1989}. In 3D systems, extremal parts of the Fermi surface
dominate the oscillations, since only at these points is the area changing slowly as a
function of $k_z$. Here, in a quasi-2D system where the warping is small compared to the
Landau level spacing, there are oscillatory contributions at all values of $k_z$, not just
from the extrema. In the simplest case of a single warping parameter this integrates to give
the Bessel function $J_0$. This avoids the need for separate amplitude and phase parameters
for the neck and belly and allows one to make the simplifying assumption of a single mass and
scattering rate for the whole surface.

 The three amplitude reduction factors, $R_T$, $R_D$, and $R_s$
account for the effects of temperature, scattering, and spin splitting, respectively.
Temperature broadens the Fermi distribution, making the impact of a Landau level crossing the
Fermi surface less dramatic as temperature increases. This factor is
\begin{equation}\label{RT}
R_T = \frac{2 \pi^2 k_B T m^*/e \hbar H\cos\theta}{\sinh \left(2 \pi^2 k_B T m^*/e \hbar H
\cos\theta \right)},
\end{equation}
 where $m^*$ is the cyclotron mass of the electron measured with the field perpendicular to the copper oxygen planes ($\theta = 0$). The $cos\theta$ terms arise because the mass
 goes as $m^*/cos\theta$ in a quasi-2d system \cite{Singleton:2000}. The relaxation time
of the electrons due to scattering widens the Landau levels in accordance with the
uncertainty principle\cite{Schoenberg:1984}. This effect is given by
\begin{equation}\label{RD}R_D = \exp\left(- H_D(\theta) / H\right),
\end{equation} where $H_D = \frac{2\pi^2m^*}{e\cos(\theta)}\frac{\overline{v}_F}{l_{free}}$ is a characteristic field that only depends on the average Fermi velocity around the orbit
 $\overline{v}_F$ and the mean free path $l_{free}$ \cite{Bergemann:2003}. For a quasi 2D metal, $v_F$ is almost constant around any orbit. This is because
  the velocity vector is always perpendicular to the Fermi surface, constraining the orbit of the electron in real space to lie largely in the conducting plane.
Some of the scattering going into $l_{free}$ likely comes from scattering due to vortices,
but we leave this as a field-independent scattering contribution, provided that the analysis
is restricted to the field range above which the resistivity becomes only weakly
field-dependent in the vortex liquid state (arrows in figure 1).

Finally, the degeneracy of each Landau level is lifted by the Zeeman effect, which splits the
energy of spin up and spin down contributions by an amount linear in the applied field (the
splitting is $g\mu_B H$ where $\mu_B$ is the Bohr magneton; a $g$ factor near 2 is the value
expected for free spins). This splitting results in interference between the oscillations on
the spin-up and spin-down Fermi surfaces. The effect is given by
\begin{equation}\label{Rs}Rs = \cos\left(\frac{\pi g m_s}{2 m_e \cos\theta}\right),
\end{equation}
where $m_s$ is the `spin' mass of the electron at $\theta = 0$, related to the cyclotron mass
by $m^* = (1 + \lambda)m_s$, where $\lambda$ is the mass enchancement due to electron-phonon
interactions\cite{Schoenberg:1984,Engelsberg:1970}. Note that while the cyclotron mass
appearing in the temperature dependent $R_T$ term is fully renormalized by both
electron-electron and electron-phonon interactions, the mass appearing in $R_s$ is
renormalized only by electron-electron interactions. This result, first shown by Fowler and
Prange\cite{Fowler:1965}, means that the product $g m_s$ is accessible through the angular
dependence of the amplitude of the oscillations \cite{Singleton:2000}. It should be noted
that this term is independent of field, so that when $g m_s/m_e \cos\theta$ is an odd
integer, the entire oscillatory signal is dampened to zero. On the other hand, the $\sin
* J_0$ term in Eq. 1, which produces an angle dependent node in the data, \textit{but
only at certain field values.}

The fact that the oscillatory amplitude does not go right to zero at the node in the field
dependence near 47 Tesla (Fig. 1 inset) suggests that there is more than one Fermi cylinder,
since the Bessel function for simple sinusoidal warping of one cylinder produces a full node
at a field where the neck and belly frequencies interfere destructively. To constrain the fit
to two cylinders, each with terms of the form given in Eq.\ref{osc}, we have made
measurements at several angles of tilt, $\theta$, of the magnetic field away from the
$\hat{c}$-axis, which probes the degree of warping of the cylinders. A standard method used
to analyze such data is to extract individual frequency components using Fourier transforms.
The shape of the Fermi surface is then mapped using the angle-dependence of the frequencies.
Fits of the amplitude to the temperature, field and angle dependence determine $R_T$, $R_D$
and $R_s$. The Fourier transform treatment does not work for underdoped $YBa_2Cu_3O_{6.59}$
because we cannot unambiguously separate the individual frequencies. Furthermore, fitting the
field dependent oscillations separately at each angle and temperature does not yield a
self-consistent set of parameters, because a single sweep can not constrain all four of the
factors affecting the amplitude. However, a fit to the field and temperature dependence at a
fixed angle can constrain $R_T$, and a fit to the field and angle dependence can constrain
$R_S$. A global fit to all data leaves only the overall amplitude and the mean free path
$l_{free}$ coupled. We present here a method for making a global fit which uses a genetic
algorithm that explores a wide parameter space for the entire data set, fitting both the
field and the angle dependence of the resistivity simultaneously and then iterating with fits
to the field and temperature dependence. Details of the analysis technique and its advantages
over Fourier transform methods are given in the supplementary material.

Fig. 2a shows the results of the field-angle fit from 1.6 to 45.3 degrees, taken at 4.2
Kelvin. The model is nearly indistinguishable from the data and any small differences could
be due to the neglect of the contributions from higher harmonics and a very weak signal with
a frequency near 660 Tesla. This weak component is not the same as the the large, extremal
orbit near 600 Tesla reported by Sebastian et al. \cite{Sebastian:2010}; we also observe a
belly orbit for a warped cylinder, but at a much lower frequency (near 515 Tesla). The weak
660 Tesla component is a much smaller contribution that we omit since it is barely above the
background noise.

An alternative to two sinusoidally warped surfaces would be a single cylinder with higher
order warping terms\cite{Bergemann:2003,Harrison:2009}, but we find that two Fermi cylinders,
each with a single warping parameter, are required to get a fit that describes the data
across all angles and temperatures. For fixed angle, the field/temperature dependence can be
fitted, as shown in Fig. 2b. Both fits are iterated to determine a unique set of parameters,
since it is difficult to determine the mass $m^* = m_s(1+\lambda)$ from the field-angle fits
of Fig. 2a and the term $gm_s/m_e$ cannot be constrained in the field-temperature fits of
Fig. 2b. A summary of the parameters is given in Table 1, which shows two sheets with
slightly different average frequencies and effective masses.

These fits indicate that within this field and temperature range, the measurements are
consistent with the Lifschitz-Kosevich expressions introduced above, even though a drastic
reorganization of the Fermi surface has occurred\cite{Sebastian:2009}. There are several
scenarios as to how these cylinders are arranged within the Brillouin zone. One developed by
Sebastian et al. is that they are distinct hole and electron Fermi
surfaces\cite{Sebastian:2010}, a possibility suggested by the fact that the warping of the
two cylinders is rather different and that this might arise if the Fermi cylinders lie in
regions of the Brillouin zone that have different c-axis hopping. Alternatively, they might
be more closely related. For instance, if these cylinders were electron pockets, then the
orthorhombicity of this material would permit somewhat different cylinders at [$\pi$,0] and
[0,$\pi$]. Also, YBCO has a double layer of CuO$_2$ planes, which could give different sized
cylinders due to bilayer-splitting \cite{Carrington:2009}.

The oscillation frequencies are not the only interesting fit parameters, as there is considerable information in the effective masses $m^*$ and $m_s$. In
particular, the fact that the spin-splitting term $R_s$ lacks any mass enhancement from
electron-phonon interactions allows one to draw conclusions about the $g$-factor and the
electron-phonon coupling $\lambda$. If the $g$-factor in high field maintains the value 2.2
that has been determined by thermodynamic and NMR measurements\cite{Walstedt:1992} on YBa$_2$Cu$_3$O$_{7}$ averaged across the
entire Fermi surface at low and moderate field, then
  one can calculate $\lambda$.
 Our fitted values give an estimate of $\lambda = .58$ and  $\lambda = .18$ for the surfaces of average area $478 T$ and $526 T$, respectively.
 However, if the value of $g$ is larger than 2.2,
 $\lambda$ could take on higher values.

\begin{table}   \label{tab:fit}
    \centering
        \begin{tabular}{| l || c | c |}
        \hline
        & Surface 1 & Surface 2\\ \hline
        $F [$T$]$ & $478$ & $526$\\ \hline
        $\Delta F [$T$]$ & $37.7$ & $3.5$\\ \hline
        $m^* /m_e$ & $1.5$ & $1.7$\\ \hline
        $g m_s /m_e$ & $2.1$ & $3.2$ \\ \hline
      $l_{free} [$\AA$]$ & $387$ & $325$ \\ \hline
      $\gamma$ & $3.5$ & $1.1$ \\ \hline
      $A$ & $13$ & $18.5$ \\ \hline

        \end{tabular}
    \caption{\textbf{Fit parameters for two warped Fermi surfaces.} The best fit parameters
obtained after iterating the resistivity-field-angle and resistivity-field-temperature fits.
Note that the ratio of $F_1$ to $F_2$ is close to the ratio of $m^*_1$ to $m^*_2$, which is expected.  A discussion about parameter uncertainties can be found in the supplementary information.}

\end{table}

A much stronger statement can be made about $g$. Since  $\lambda$ is positive, by taking the
electron-phonon coupling to be zero we put an extreme lower bound on $g$ of $1.39$ and $1.86$
for the smaller and larger surfaces, respectively. The spin-splitting term $R_s$ is unique
among the amplitude reduction factors $R_s$, $R_T$, and $R_D$ in that it is non-monotonic in
its angle dependence; for a single Fermi surface, the $R_s$ term drives the oscillation
amplitude at all fields to zero at a particular angle, followed by a recovery and a
$\pi$-phase shift. A recent search for such a ``spin-zero'' did not observe this
effect\cite{Sebastian:2009b}. However, with more than one Fermi surface, the spin-zeros of
each can be at different angles
 and the effect is made less obvious. In order to better display the important role that spin-splitting plays
 we have made a series of individual fits up to higher angles holding all parameters
 from Table 1 fixed but allowing the amplitude parameter Rs to vary independently at each angle. The higher angle data was not included in the earlier fits
because at the highest angles one must make use of the weak oscillations seen just as the
vortex lattice is melting, the regime of rapid change in resistivity seen in figure 1 between
20 and 30 Tesla. Figure 3a shows that for each of the Fermi cylinders there is an angle at
which the amplitude falls to zero, followed by a recovery. The impact of this non-monotonic
angle dependence is apparent in the raw data in two regimes. Even at low angles in Fig. 2a,
the amplitude stays large or even increases somewhat with increasing angle, even though $R_T$
and $R_D$ should be driving the amplitude downwards. At high angles, Fig. 3b shows a rapid
drop in amplitude as the spin zero is approached, followed by a roughly constant amplitude as
the $R_S$ competes with the overall trend of the other amplitude terms. The phase flip upon
passing through the spin zero is not unambiguously observed until both cylinders have passed
through their spin zero.

Figure 4 highlights the role of Zeeman splitting of the spin states. While there are
angle-dependent nodes due to the warping of the two Fermi cylinders, there is a key feature
that is independent of field. Namely, the amplitude exhibits a minimum for all fields at
around $\theta= 50$ degrees, due to the interference of spin-up and spin-down contributions
on each Fermi sheet. Both sheets have a value of $g$ that is far from zero and that is close
to the value of $g$ for free electrons. This rules out a number of reconstruction scenarios
involving spin order, particularly a conventional spin density wave with staggered moments
parallel to the $CuO_2$ planes \cite{Norman:2010,Garcia:2010,Ramazashvili:2010}.

methods: The crystals used in this study were produced using the same self-flux growth
technique as our previous studies, but with a recent improvement in purity by employing
99.9997\% pure CuO$_2$ in the starting materials. Oxygen concentration has been set to
YBa$_2$Cu$_3$O$_{6.59}$ by annealing in controlled oxygen partial pressure, followed by a
homogenization stage in which the crystals are sealed with a large volume of ceramic at the
same oxygen content so that they can be slowly heated and cooled in equilibrium. The samples
are mechanically detwinned by heating under uniaxial stress and $\hat{c}$-axis resistivity
contacts are applied in a Corbino-like geometry by evaporating gold through masks. The
relatively high resistance along the $\hat{c}$-axis carries the advantage that for fixed
current one can obtain very high signal-to-noise, providing cleaner data with no need for
averaging.

  \textbf{Acknowledgements} The authors would like to thank S. Julian, R. Ramazashvili, L. Taillefer, S. Kivelson, S. Sebastian, G. Lonzarich, M. Berciu, S. Sachdev, S. Chakravarty,
 P.C.E Stamp, L. Thompson and M. Norman for many helpful conversations.
Research support was provided by the Canadian Institute for Advanced
Research, the Natural Science and Engineering Research Council, the French
ANR DELICE, and Euromagnet II

  \textbf{Competing Interests} The authors declare that they have no
competing financial interests.

 \textbf{Correspondence} Correspondence and requests for materials
should be addressed to D.~A.~Bonn~(email: bonn@phas.ubc.ca).

\bibliography{bradrefs}

\begin{thebibliography}{10}
\expandafter\ifx\csname url\endcsname\relax
  \def\url#1{\texttt{#1}}\fi
\expandafter\ifx\csname urlprefix\endcsname\relax\def\urlprefix{URL }\fi
\providecommand{\bibinfo}[2]{#2}
\providecommand{\eprint}[2][]{\url{#2}}

\bibitem{Doiron:2007}
\bibinfo{author}{Doiron-Leyraud, N.} \emph{et~al.}
\newblock \bibinfo{title}{Quantum oscillations and the {Fermi} surface in an
  underdoped high {Tc} superconductor}.
\newblock \emph{\bibinfo{journal}{Nature}} \textbf{\bibinfo{volume}{447}},
  \bibinfo{pages}{565} (\bibinfo{year}{2007}).

\bibitem{Jaudet:2008}
\bibinfo{author}{Jaudet, C.} \emph{et~al.}
\newblock \bibinfo{title}{{de Haas-van Alphen} oscillations in the underdoped
  high-temperature superconductor {YBa$_2$Cu$_3$O$_{6.5}$}}.
\newblock \emph{\bibinfo{journal}{Phys. Rev. Lett.}}
  \textbf{\bibinfo{volume}{100}}, \bibinfo{pages}{187005}
  (\bibinfo{year}{2008}).

\bibitem{Sebastian:2008}
\bibinfo{author}{Sebastian, S.} \emph{et~al.}
\newblock \bibinfo{title}{Multi-component {Fermi} surface in an underdoped high
  temperature superconductor}.
\newblock \emph{\bibinfo{journal}{Nature}} \textbf{\bibinfo{volume}{454}},
  \bibinfo{pages}{200} (\bibinfo{year}{2008}).

\bibitem{Bangura:2008}
\bibinfo{author}{Bangura, A.} \emph{et~al.}
\newblock \bibinfo{title}{Small {Fermi} surface pockets in underdoped high
  temperature superconductors: Observation of {Shubnikov-de Haas} oscillations
  in {YBa$_2$Cu$_4$O$_8$}}.
\newblock \emph{\bibinfo{journal}{Phys. Rev. Lett.}}
  \textbf{\bibinfo{volume}{100}}, \bibinfo{pages}{047004}
  (\bibinfo{year}{2008}).

\bibitem{Yelland:2008}
\bibinfo{author}{Yelland, E.} \emph{et~al.}
\newblock \bibinfo{title}{Quantum oscillations in the underdoped cuprate
  {YBa$_2$Cu$_4$O$_8$}}.
\newblock \emph{\bibinfo{journal}{Phys. Rev. Lett.}}
  \textbf{\bibinfo{volume}{100}}, \bibinfo{pages}{047003}
  (\bibinfo{year}{2008}).

\bibitem{Vignolle:2008}
\bibinfo{author}{Vignolle, B.} \emph{et~al.}
\newblock \bibinfo{title}{Quantum oscillations in an overdoped high-{Tc}
  superconductor}.
\newblock \emph{\bibinfo{journal}{Nature}} \textbf{\bibinfo{volume}{455}},
  \bibinfo{pages}{952} (\bibinfo{year}{2008}).

\bibitem{Helm:2009}
\bibinfo{author}{Helm, T.} \emph{et~al.}
\newblock \bibinfo{title}{Evolution of the fermi surface of the electron-doped
  high-temperature superconductor {Nd$_{2-x}$Ce$_x$CuO$_4$} revealed by
  {Shubnikov-de Haas} oscillations}.
\newblock \emph{\bibinfo{journal}{Phys. Rev. Lett.}}
  \textbf{\bibinfo{volume}{103}}, \bibinfo{pages}{157002}
  (\bibinfo{year}{2008}).

\bibitem{Audouard:2009}
\bibinfo{author}{Audouard, A.} \emph{et~al.}
\newblock \bibinfo{title}{Multiple frequencies in quantum oscillations from
  underdoped {$YBa_2Cu_3O_{6.5}$}}.
\newblock \emph{\bibinfo{journal}{Phys. Rev. Lett.}}
  \textbf{\bibinfo{volume}{103}}, \bibinfo{pages}{157003}
  (\bibinfo{year}{2009}).

\bibitem{Sebastian:2010}
\bibinfo{author}{Sebastian, S.} \emph{et~al.}
\newblock \bibinfo{title}{Compensated electron and hole pockets in an
  underdoped high {Tc} superconductor}.
\newblock \emph{\bibinfo{journal}{Phys. Rev. B}} \textbf{\bibinfo{volume}{81}},
  \bibinfo{pages}{214524} (\bibinfo{year}{2010}).

\bibitem{Bergemann:2003}
\bibinfo{author}{Bergemann, C.} \emph{et~al.}
\newblock \bibinfo{title}{Quasi-two-dimensional {Fermi} liquid properties of
  the unconventional superconductor {Sr$_2$RuO$_4$}}.
\newblock \emph{\bibinfo{journal}{Advances in Physics}}
  \textbf{\bibinfo{volume}{52}}, \bibinfo{pages}{639} (\bibinfo{year}{2003}).

\bibitem{Bergemann:2002}
\bibinfo{author}{Bergemann, C.} \emph{et~al.}
\newblock \bibinfo{title}{Detailed topography of the {Fermi} surface of
  {Sr$_2$RuO$_4$}}.
\newblock \emph{\bibinfo{journal}{Phys. Rev. Lett.}}
  \textbf{\bibinfo{volume}{84}}, \bibinfo{pages}{2662} (\bibinfo{year}{2000}).

\bibitem{Yamaji:1989}
\bibinfo{author}{Yamaji, K.}
\newblock \bibinfo{title}{On the angle dependence of the magnetoresistance in
  quasi-two-dimensional organic superconductors}.
\newblock \emph{\bibinfo{journal}{Journal of the Physical Society of Japan}}
  \textbf{\bibinfo{volume}{58}}, \bibinfo{pages}{1520} (\bibinfo{year}{1989}).

\bibitem{Singleton:2000}
\bibinfo{author}{Singleton, J.}
\newblock \bibinfo{title}{Studies of quasi-two-dimensional organic conductors
  based on {BEDT-TTF} using high magnetic fields}.
\newblock \emph{\bibinfo{journal}{Rep. Prog. Phys.}}
  \textbf{\bibinfo{volume}{63}}, \bibinfo{pages}{1111} (\bibinfo{year}{2000}).

\bibitem{Schoenberg:1984}
\bibinfo{author}{Schoenberg, D.}
\newblock \emph{\bibinfo{title}{Magnetic Oscillations in Metals}}
  (\bibinfo{publisher}{Cambridge University Press},
  \bibinfo{address}{Cambridge}, \bibinfo{year}{1984}).

\bibitem{Engelsberg:1970}
\bibinfo{author}{Engelsberg, S.} \emph{et~al.}
\newblock \bibinfo{title}{Influence of electron-phonon interactions on the {de
  Haas-van Alphen} effect}.
\newblock \emph{\bibinfo{journal}{Physical Review B}}
  \textbf{\bibinfo{volume}{2}}, \bibinfo{pages}{1657} (\bibinfo{year}{1970}).

\bibitem{Fowler:1965}
\bibinfo{author}{Fowler, M.} \emph{et~al.}
\newblock \bibinfo{title}{Electron-phonon renormalization effects in high
  magnetic field. the {de Haas-van Alphen} effect}.
\newblock \emph{\bibinfo{journal}{Physics}} \textbf{\bibinfo{volume}{1}},
  \bibinfo{pages}{315} (\bibinfo{year}{1965}).

\bibitem{Harrison:2009}
\bibinfo{author}{Harrison, N.} \emph{et~al.}
\newblock \bibinfo{title}{Determining the in-plane {Fermi} surface topology in
  underdoped high {Tc} superconductors using angle-dependent magnetic quantum
  oscillations}.
\newblock \emph{\bibinfo{journal}{Journal of Physics: Condensed Matter}}
  \textbf{\bibinfo{volume}{21}}, \bibinfo{pages}{1} (\bibinfo{year}{2009}).

\bibitem{Sebastian:2009}
\bibinfo{author}{Sebastian, S.} \emph{et~al.}
\newblock \bibinfo{title}{Fermi liquid behaviour in an underdoped high {Tc}
  superconductor}.
\newblock \emph{\bibinfo{journal}{arXiv:0912.3022}}  (\bibinfo{year}{2009}).

\bibitem{Carrington:2009}
\bibinfo{author}{Carrington, A.} \emph{et~al.}
\newblock \bibinfo{title}{Quantum oscillation studies of the fermi surface of
  {LaFePO}}.
\newblock \emph{\bibinfo{journal}{Physica C}} \textbf{\bibinfo{volume}{469}}
  (\bibinfo{year}{2009}).

\bibitem{Walstedt:1992}
\bibinfo{author}{Walstedt, R.} \emph{et~al.}
\newblock \bibinfo{title}{Diamagnetism in the normal state of {YBCO}}.
\newblock \emph{\bibinfo{journal}{Phys. Rev. B}} \textbf{\bibinfo{volume}{45}},
  \bibinfo{pages}{8074} (\bibinfo{year}{1992}).

\bibitem{Sebastian:2009b}
\bibinfo{author}{Sebastian, S.} \emph{et~al.}
\newblock \bibinfo{title}{Spin-order driven {Fermi} surface reconstruction
  revealed by quantum oscillations in an underdoped high {Tc} superconductor}.
\newblock \emph{\bibinfo{journal}{Phys. Rev. Lett.}}
  \textbf{\bibinfo{volume}{103}}, \bibinfo{pages}{256405}
  (\bibinfo{year}{2009}).

\bibitem{Norman:2010}
\bibinfo{author}{Norman, M.~R.} \emph{et~al.}
\newblock \bibinfo{title}{Spin zeros and the origin of fermi surface
  reconstruction in the cuprates}.
\newblock \emph{\bibinfo{journal}{arXiv:1007.1047v1}}  (\bibinfo{year}{2010}).

\bibitem{Garcia:2010}
\bibinfo{author}{Garcia-Aldea, D.} \emph{et~al.}
\newblock \bibinfo{title}{Singlet versus triplet particle-hole condensates in
  quantum oscillations in cuprates}.
\newblock \emph{\bibinfo{journal}{arXiv:1008.2030v1}}  (\bibinfo{year}{2010}).

\bibitem{Ramazashvili:2010}
\bibinfo{author}{Ramazashvili, R.}
\newblock \bibinfo{title}{Quantum oscillations in antiferromagnetic conductors
  with small carrier pockets}.
\newblock \emph{\bibinfo{journal}{arXiv:1006.0167v1}}  (\bibinfo{year}{2010}).

\bibitem{Harris:1978}
\bibinfo{author}{Harris, F.~J.}
\newblock \bibinfo{title}{On the use of windows for harmonic analysis with the
  discrete fourier transform}.
\newblock \emph{\bibinfo{journal}{Proc IEEE}} \textbf{\bibinfo{volume}{66}},
  \bibinfo{pages}{51} (\bibinfo{year}{1978}).

\bibitem{Benzhaf:1998}
\bibinfo{author}{Banzhaf, W.}, \bibinfo{author}{Nordin, P.},
  \bibinfo{author}{Keller, R.} \& \bibinfo{author}{Francone, F.}
\newblock \emph{\bibinfo{title}{Genetic Programming - An Introduction}}
  (\bibinfo{publisher}{Morgan Kaufmann, San Francisco, CA.},
  \bibinfo{year}{1988}).

\end{thebibliography}

\begin{center}
\textbf{Supplementary Information}
\end{center}
\section{Fourier Transform Limitations}

In order to illustrate the difficulties encountered in using Fourier transforms for quantum
oscillation data in the cuprates, a comparison can be made with the more favourable situation in
the ruthenates.\cite{Bergemann:2002} The smallest piece of Fermi surface in Sr$_2$RuO$_4$,
the alpha pocket, has a frequency of 2.6 kT. Because the T$_c$ is so low, 1.43 Kelvin, the
oscillations are visible at very low fields, and there are about 800 visible oscillations
between 3 and 33 Tesla; the larger Fermi pockets have several thousand oscillations in this
same range. When Fourier transforming a finite number of oscillations, the resulting
frequency spectrum is a convolution of the Fourier transform of the oscillating signal, which
is a collection of narrow peaks, with the Fourier transform of the boxcar function (which is
the sinc function) corresponding to the upper and lower limit of the data. If the number of
oscillations within the box is small, then one gets large side lobes and considerable
broadening of each peak in the frequency spectrum. Additionally, because the amplitude of the
oscillations is field dependent, the approximately exponential envelope results in a further
broadening of the Fourier peaks. In the case of Sr$_2$RuO$_4$, Bergemann et
al.\cite{Bergemann:2003} used Fourier transforms over a window that contains both a large
number of oscillations yet a relatively constant amplitude. This fact, coupled with the
additional advantage that the oscillation frequencies are well spaced out in frequency space,
results in well defined and well separated peaks for each piece of Fermi surface.

For YBa$_2$Cu$_3$O$_{6+x}$, the small number of oscillations, the strongly varying background
over the field window, and the fact that the frequencies from the two pockets are close
together result in broad features in frequency space that overlap each other. Because the
Fourier transform gives a complex number, the results are traditionally displayed as the
absolute value of this number - a power spectrum. Frequencies that are close together
interfere with each other, resulting in relative amplitudes and peak frequencies in Fourier
space that do not always reflect the amplitudes and frequencies of the oscillating components
in the original data. There are a number of other windows that can be used to apodize the
data (other than the boxcar function) in order to decrease the side lobe amplitudes, each
having their own tradeoffs (more sidelobes for a narrower central peak, fewer sidelobes for a
broader central peak, and mixtures of the two)\cite{Harris:1978}. None of these address the phase interference
of closely spaced frequencies. To quantify the problem, the frequency resolution of a power
spectrum is limited by the number of oscillations measured. With a field range of 30 to 60
Tesla, one is limited to a resolution of 1/(1/30-1/60) = 60 Tesla. This is larger than the
difference between the average areas of the two pieces of Fermi surface in
YBa$_2$Cu$_3$O$_{6+x}$. Going to 80 Tesla only increases the resolution to 48 Tesla because
the oscillations are periodic in 1/B, so one only gains 2.2 periods of oscillation. If the
exact shape of the amplitude envelope and windowing function were known, and there was no
problem with background subtraction, the peaks in Fourier space could be fit to a model and
the frequencies extracted with much higher precision. However, the exact shape of the
background for each surface is not known a priori, so one would still need to do at least
some fitting to the original data just to determine the background to be subtracted.

\section{Genetic Algorithm}

Given the difficulties noted above, a direct fit to the raw data is preferable, and can
provide highly constrained parameters if the signal to noise ratio is high. In the fits
presented in this article, even the simplest model has many parameters and must be fit to the
entire data set. A very useful tool for this is a genetic search algorithm inspired by
evolution as it occurs in nature, taking a heuristic approach to finding the optimal solution
rather than a deterministic one\cite{Benzhaf:1998}. The problem is finding the best set of
parameters for a model given a certain set of data points, and the optimal solution is the
set of parameters that gives the smallest least squares value. A gradient based search
algorithm such as Newton's method may become stuck in local minima if the parameter "landscape" is
very "rough", making the fit obtained very sensitive to the initial conditions chosen. The
genetic algorithm does not take a user-specified starting point, but rather a range for the
parameters in which the solution is thought to lie, and then the algorithm randomly generates
a large number of starting points. Local minima can be "escaped" by virtue of the design of
the algorithm, and the propensity of the algorithm to search outside of the minimum to which it is
converging is a controllable parameter. The algorithm proceeds as follows:

1. A large number (N) of initial parameter sets (called $x_{parent}$) are created,about 10
times as many sets as there are parameters. In our case the parameters were the average
frequency, the warping size, the Dingle factor, an amplitude factor, a phase factor, and
$g*m_s$ (the spin susceptibility); there were two of each of these sets - one for each piece
of Fermi surface. Bounds are specified for the initial values: for example, the average size
of the first cylinder could be restricted to lie between 400 and 700 Tesla.

2. A second new parameter set (the "mutated" set) is created by taking three sets at random
and creating a new set via $x_{mut} = x_k  + s (x_l - x_m )$, where $s$ is a user defined
scaling factor. Note that each $x_i$ is a vector of parameters, so the algebra is performed
on each parameter within the vector. If any parameter ends up outside its original specified
bounds, then its value is set to the boundary value. This is repeated until there are N new
"mutated" parameter sets.

3. The "crossing" is then performed by taking the $j^{th}$ set from the parent parameter sets
and the $j^{th}$ set from the mutated parameter sets. Starting with the first parameter (say
the average frequency of the first piece of Fermi surface), a new parameter set $x_{new}$  is
created by selecting the first parameter from the original set with probability $p$, and from
the mutated set with probability $1-p$. This is repeated for each parameter in the set until
$x_{new}$ is completed.  This is then repeated for all N parameter sets.

4. Competition and selection is then effected by computing the sum of the squares of differences
between the model with the data points using the parameters from both the $j^{th}$ set from
$x_{new}$ and the $j^{th}$ set from $x_{parent}$ and then keeping the parameter set that has
the lowest sum of squares. This is repeated for each of the N pairs.

5. If the lowest sum of the squares value is less than the tolerance specified, the algorithm
terminates, otherwise it returns to step 2.

This algorithm is still capable of becoming stuck in local minima because the initial population is finite,
 and so was run about a hundred times. Because each run is independent from the others, this can be done in parallel,
reducing the computation time. Various adjustments can be made to the algorithm, including
(but not limited to): always keeping a few of the best parameter sets from each generation
(called "elitism"), penalizing parameter sets that are sitting on the boundary (effectively
increasing their least squares value in some artificial way), or even changing the order of
some of the steps.

It should be noted that the statistical cuncertainties that have been caculated for the fit parameters have not been included in table 4
because the authors feel that they are much smaller than the uncertainty in the model itself.  For example, the largely 
unwarped cylinder has an average frequency of 526 Tesla has, within the model we used, a statistical uncertainty of $+/-  0.8$ Tesla. Including 
higher warping harmonics on the other warped cylinder, adding more Fermi cylinders, accounting for oscillations of the chemical potential itself, including 
scattering between adjacent Landau levels, and a host of other subtle effects could all change this number by an amount greater than $0.8$ Tesla.  That being said,
we beleive that the simplest possible model that reproduces all the features of the data to within the noise of the data is the best starting point.

\newpage

\begin{figure}
\includegraphics{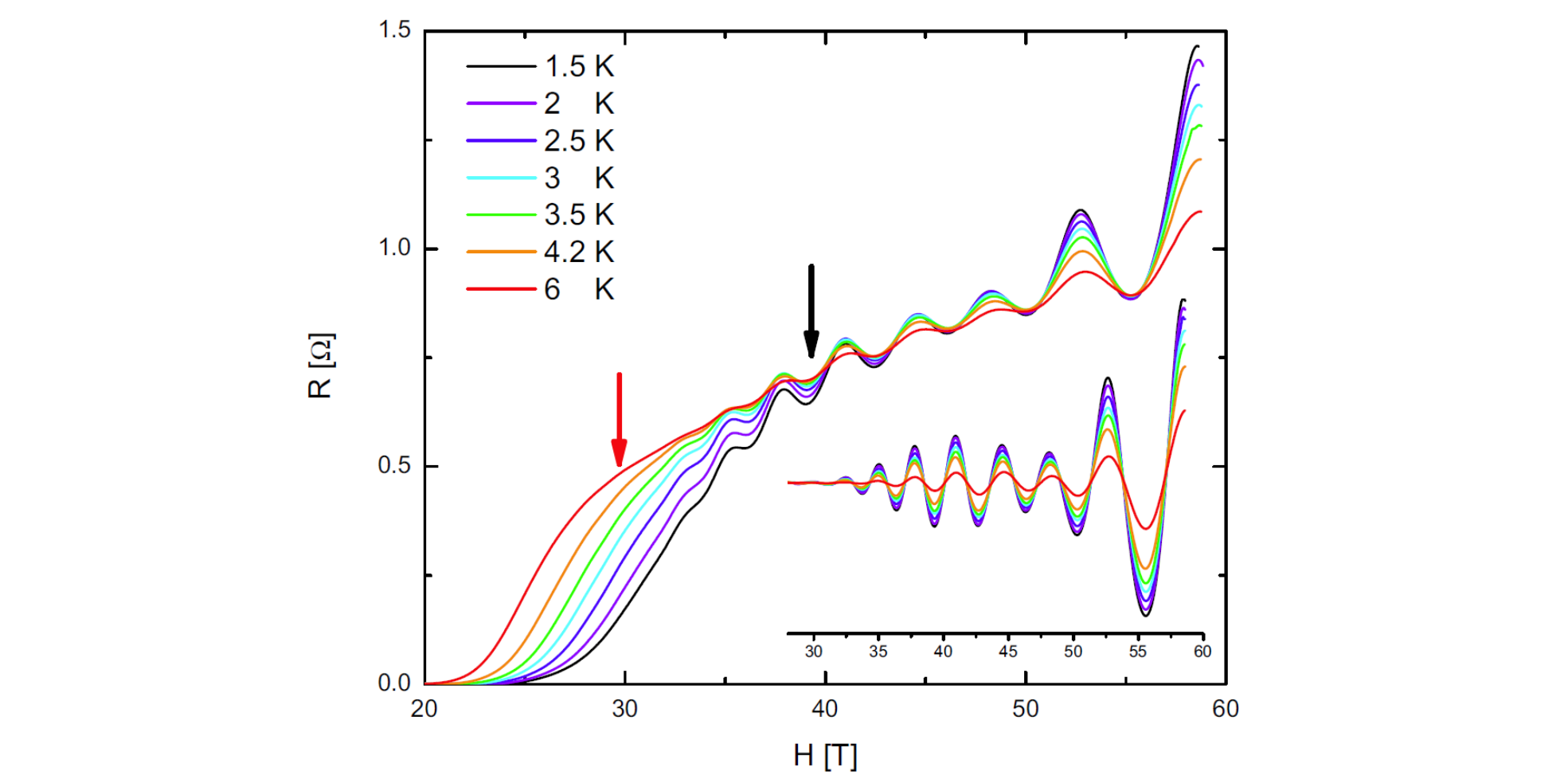}
\caption{\textbf{Field Dependence of the $\hat{c}$-axis resistance of
YBa$_\mathbf{2}$Cu$_\mathbf{3}$O$_\mathbf{6.59}$.}  $\hat{c}$-axis resistance obtained during
the down sweep of an 80 millisecond field pulse with the field parallel to the
$\hat{c}$-axis.   The black and red arrows indicate the minimum field used in the temperature
dependence fits, for 1.5 and 6 Kelvin, respectively.  Below these field values, the
scattering is stronger and field dependent, possibly due to the melting of the vortex
lattice. } \label{fig:rawdata}
\end{figure}

\newpage

\begin{figure}
\includegraphics{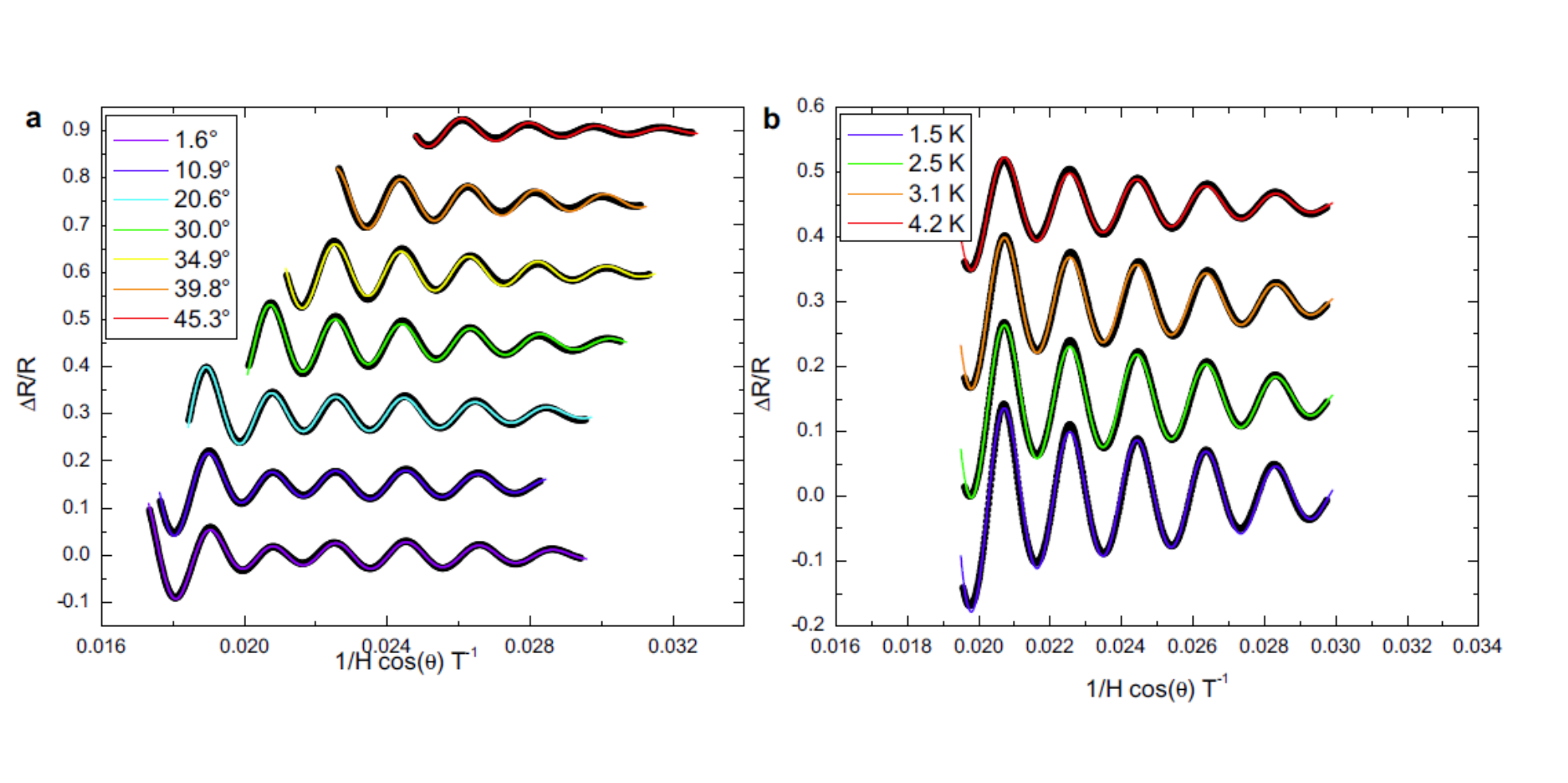}
\caption{\textbf{Fits to the angle and temperature dependence of the oscillatory component}
\textbf{a.} $\Delta R/R$ at 4.2 Kelvin as a function of field, with the field tilted from 1.6
to 45.3 degrees with respect to the c-axis. The offset of the crystal allignment from
0 degrees was checked by measuring at negative angles (not shown). The data are in black and the fits are in colour.
The data have been vertically offset for clarity. As the average frequency $ F_0$ is expected
to scale as $1/\cos\theta$, these data sets are plotted versus $1/H\cos\theta$.  The fits to
the angle dependence give the product of the $g$ factor with the band mass renormalized only
by electron-electron interactions: $gm_{s1} = 2.12 m_e$ and $gm_{s1} = 3.18 m_e$. \textbf{b.}
$\Delta R/R$ as a function of field at 1.5, 2.5, 3.1, and 4.2 Kelvin and with the field at
-27.9 degrees from the $c$-axis.  The data are in black and the fits are in colour, again
plotted versus $1/H\cos\theta$. Because the cyclotron mass has a simple $1/\cos\theta$
dependence, it can be extracted from the temperature dependence at any angle.  The values
obtained for the two surfaces are $m^*_{1} = 1.56 m_e$ and $m^*_{2} = 1.70 m_e$, checked
against the temperature dependence at $\theta = 1.4$ degrees for consistency. }
\label{fig:AngleTemperature}
\end{figure}

\newpage

\begin{figure}
\includegraphics{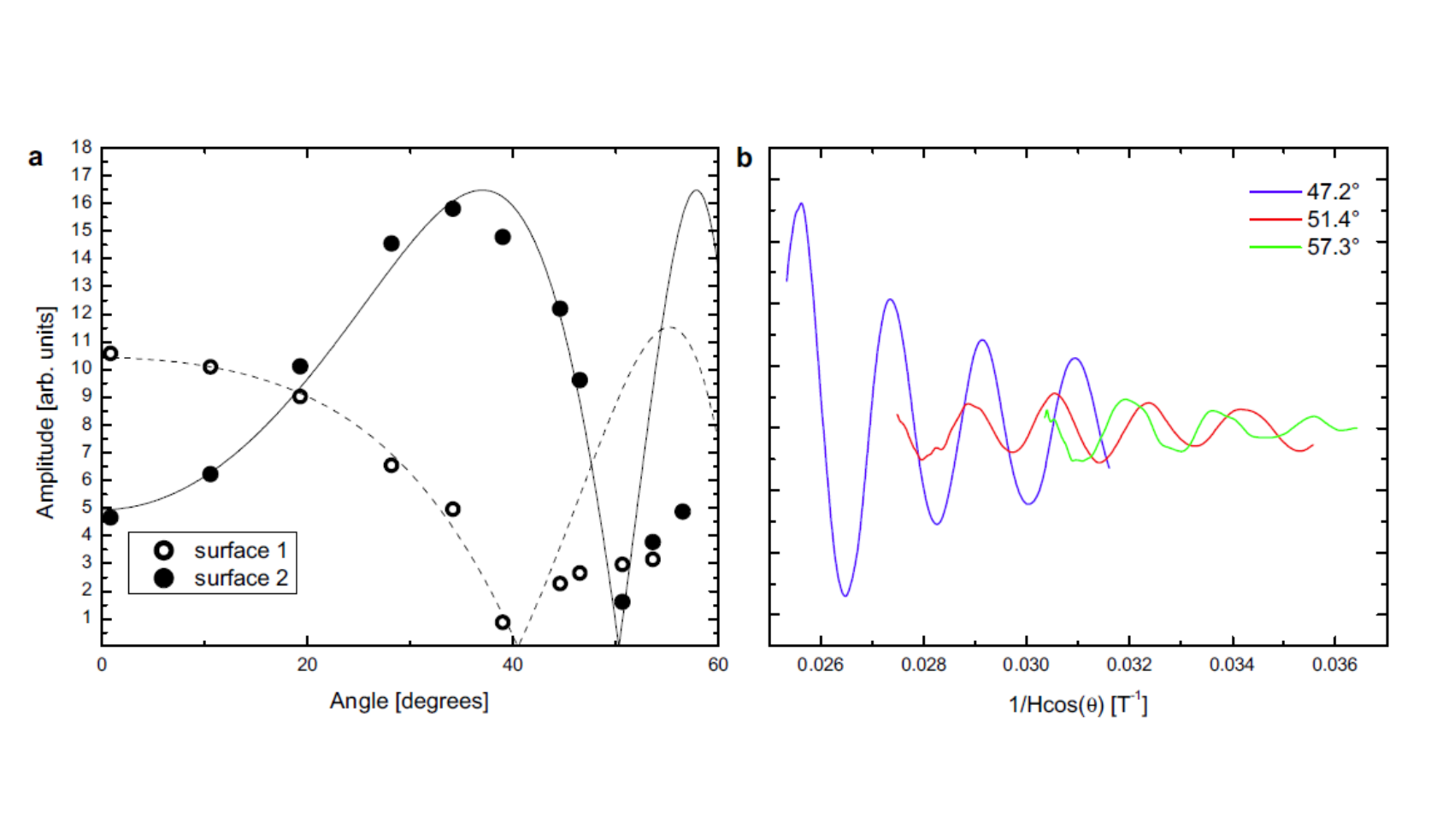}
\caption{\textbf{ ({\bf a.})Oscillatory amplitude as a function of angle from the $\hat{c}$-axis}  The
oscillation amplitude of each Fermi surface, designated ``Surface 1" and ``Surface
2" in Table 1, as a function of angle after dividing out the monotonic angle
dependence due to the Dingle factor $R_D$ and the Lifshitz-Kosevich term $R_T$,
leaving only the angle dependence due to the Zeeman spin-splitting of the Fermi surface.
The data are the black circles and the fits to the data are the lines.  At the highest
angles, the data fall below the fit line. This is likely due to the large scattering rates in
the melting transition, the region from which these data were collected, which serve to
further suppress the amplitude predicted by LK. Nonetheless, a minimum in the amplitude is
clearly seen around the fitted value of the spin zero, with the amplitude recovering at
higher angles. ({\bf b.}) Raw data used to extract the amplitudes of each surface at angles
higher than the global fit.  } \label{fig:SpinZeros}
\end{figure}

\newpage

\begin{figure}
\includegraphics{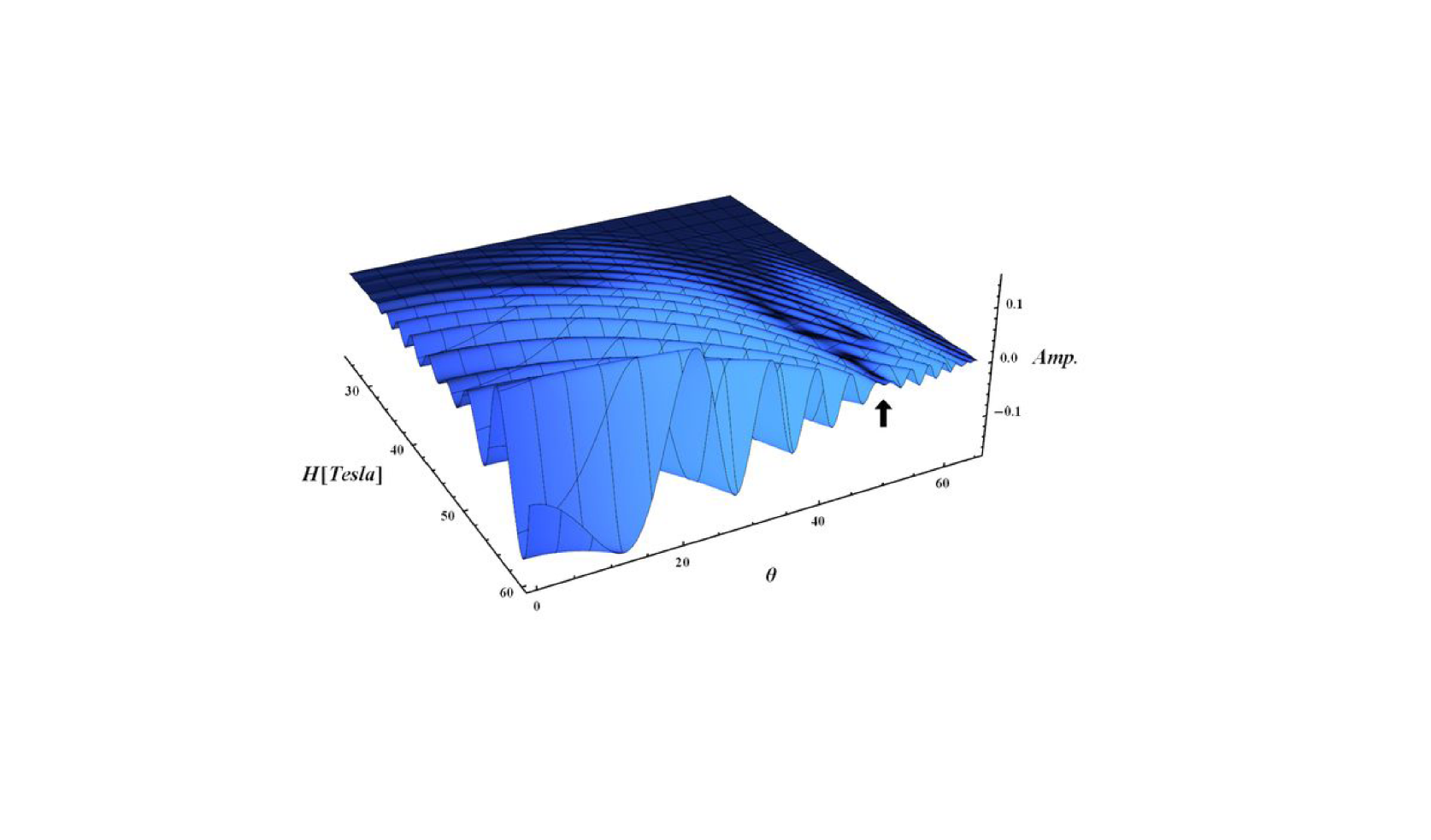}
\caption{\textbf{Oscillatory component versus angle and field} A 3D plot of the fit to the
data, showing the simultaneous evolution of the oscillations in both field and angle.  A
strong suppression of the amplitude near 50 degrees is seen, but the amplitude never goes
quite to zero due to the spin zeroes occurring at different angles (51 degrees for one, 43
degrees for the other). } \label{fig:3D}
\end{figure}

\newpage

\end{document}